\documentclass{ieeeaccess}
\usepackage{cite}
\usepackage{amsmath,amssymb,amsfonts}
\usepackage{algorithmic}
\usepackage{graphicx}
\usepackage{textcomp}

\usepackage{bm}
\makeatletter
\AtBeginDocument{\DeclareMathVersion{bold}
\SetSymbolFont{operators}{bold}{T1}{times}{b}{n}
\SetSymbolFont{NewLetters}{bold}{T1}{times}{b}{it}
\SetMathAlphabet{\mathrm}{bold}{T1}{times}{b}{n}
\SetMathAlphabet{\mathit}{bold}{T1}{times}{b}{it}
\SetMathAlphabet{\mathbf}{bold}{T1}{times}{b}{n}
\SetMathAlphabet{\mathtt}{bold}{OT1}{pcr}{b}{n}
\SetSymbolFont{symbols}{bold}{OMS}{cmsy}{b}{n}
\renewcommand\boldmath{\@nomath\boldmath\mathversion{bold}}}
\makeatother

\def\BibTeX{{\rm B\kern-.05em{\sc i\kern-.025em b}\kern-.08em
    T\kern-.1667em\lower.7ex\hbox{E}\kern-.125emX}}

\usepackage{acronym}
\usepackage{booktabs}
\usepackage{makecell}
\usepackage{cleveref}
\usepackage{url}
\crefname{figure}{Fig.}{Figs.}
\Crefname{figure}{Figure}{Figures}
\crefname{table}{Table}{Tables}
\Crefname{table}{Table}{Tables}
\crefname{section}{Sect.}{Sects.}
\Crefname{section}{Section}{Sections}
\crefname{equation}{Eq.}{Eqs.}
\Crefname{equation}{Equation}{Equations}
\begin{document}

\acrodef{asd}[ASD]{anomalous sound detection}
\acrodef{mac}[MAC]{multiply-accumulate operation}
\acrodef{scac}[SCAC]{Sub-cluster AdaCos}
\acrodef{ssl}[SSL]{self-supervised learning}
\acrodef{dcase}[DCASE]{Detection and Classification of Acoustic Scenes and Events}
\acrodef{dft}[DFT]{discrete Fourier transform}
\acrodef{stft}[STFT]{short-time Fourier transform}
\acrodef{ae}[AE]{autoencoder}
\acrodef{roc}[ROC]{receiver operating characteristic}
\acrodef{auc}[AUC]{area under the \ac{roc} curve}
\acrodef{pauc}[pAUC]{partial \ac{auc}}
\acrodef{knn}[kNN]{$k$-nearest neighbor}
\acrodef{lora}[LoRA]{low-rank adaptation}
\acrodef{sota}[SOTA]{state-of-the-art}
\acrodef{pca}[PCA]{principal component analysis}
\acrodef{gevd}[GEVD]{generalized eigenvalue decomposition}
\acrodef{nrft}[NRFT]{noise-robust feature transformation}

\history{Received 1 September 2026, accepted 5 November 2026.
  Date of publication November 2026.}
\doi{10.1109/ACCESS.2026.3752607}
\vol{14}
\year{2026}

\title{Pseudo-label distillation for discriminative anomalous sound detection}
\author{\uppercase{Takuya Fujimura}\authorrefmark{1} AND \uppercase{Tomoki Toda}\authorrefmark{1}, \IEEEmembership{Senior Member, IEEE}}

\address[1]{Nagoya University, Nagoya 464-8601, Japan}

\tfootnote{This work was partly supported by JSPS KAKENHI Grant Number JP25KJ1439.}

\markboth
{T. Fujimura \headeretal: Pseudo-label distillation for discriminative anomalous sound detection}
{T. Fujimura \headeretal: Pseudo-label distillation for discriminative anomalous sound detection}

\corresp{Corresponding author: Takuya Fujimura (e-mail: fujimura.takuya@g.sp.m.is.nagoya-u.ac.jp).}

\begin{abstract}
Discriminative \ac{asd} methods train a feature extractor through a classification task using machine-information labels.
They then detect anomalies in the resulting feature space based on distances to normal samples.
The discriminative feature space effectively captures machine characteristics, leading to high \ac{asd} performance.
However, this approach benefits from detailed labels, which are costly to obtain.
An alternative is a \ac{ssl}-based label-free approach.
This approach directly uses \ac{ssl} features for \ac{asd} and has shown competitive performance.
However, \ac{ssl} models are typically large and computationally expensive.
To address these problems, we propose a simple pseudo-label distillation framework.
The proposed method generates pseudo labels from \ac{ssl} features and trains a compact discriminative feature extractor using these pseudo labels.
To suppress the effect of noise on pseudo-label generation, we also propose lightweight \ac{nrft} methods utilizing a small amount of clean machine-sound data or isolated noise data.
We conducted comprehensive evaluations and analyses on the DCASE 2020--2025 Task~2 datasets using four \ac{ssl} models.
The results demonstrate that pseudo-label distillation not only transfers the performance of \ac{ssl} models to a compact model but also further improves performance by leveraging available coarse labels and data augmentation.
Also, our \ac{nrft} methods provide further gains.
\end{abstract}


\titlepgskip=-21pt

\maketitle

\acresetall

\section{Introduction}
\label{sec:introduction}
\Ac{asd} plays an important role in machine condition monitoring by automatically detecting mechanical failures from acoustic signals emitted by machines~\cite{koizumi2020description,kawaguchi2021description,dohi2022description,dohi2023description,nishida2024description,nishida2025description}.
Since it is difficult to collect rare and diverse anomalous sounds, the \ac{asd} task is typically formulated in an unsupervised setting, in which only normal sounds are available for system development.

\Ac{asd} systems typically consist of a frontend and a backend~\cite{fujimura2025asdkit}.
The frontend first extracts features from input machine sounds.
The backend then computes an anomaly score based on the nearest-neighbor distance between a test sample and the normal training samples in the feature space.
In this framework, anomalous sounds are expected to be distant from the normal sounds in the feature space and yield high anomaly scores.
Therefore, it is crucial to construct a feature space that effectively captures the differences between normal and anomalous sounds.

There are two representative frontend approaches: discriminative learning~\cite{lopez2020speaker,giri2020self,primus2020anomalous,wilkinghoff2023design,wilkinghoff2024self,kuroyanagi2025serial,liu2022anomalous} and \ac{ssl}~\cite{saengthong2025deep,saengthong2026sub,wilkinghoff2026temporal,wilkinghoff2023on,jiang2024anopatch,zheng2024improving,han2025exploring}.
The discriminative approach uses machine-information labels, such as machine type (e.g., gearbox and bearing) and operating parameters (e.g., voltage and rotation speed).
This approach trains a classification model on normal data using these labels, and the intermediate discriminative features are then used in the backend.
These discriminative features effectively capture rich machine-specific information and lead to high \ac{asd} performance~\cite{wilkinghoff2023design,wilkinghoff2024self,kuroyanagi2025serial}.
However, there is a clear trade-off between performance and annotation cost~\cite{fujimura2025improvements,wilkinghoff2023why}.
For example, when each machine has its own microphone, as is often the case, machine-type labels can be easily obtained from the microphone index.
However, such coarse labels are insufficient for learning an effective feature space.
Detailed labels, such as operating parameters, can greatly improve performance~\cite{fujimura2025improvements,wilkinghoff2023why}, but collecting them requires additional monitoring systems or costly human annotation.
\begin{figure}[t]
    \centering
    \includegraphics[width=\linewidth]{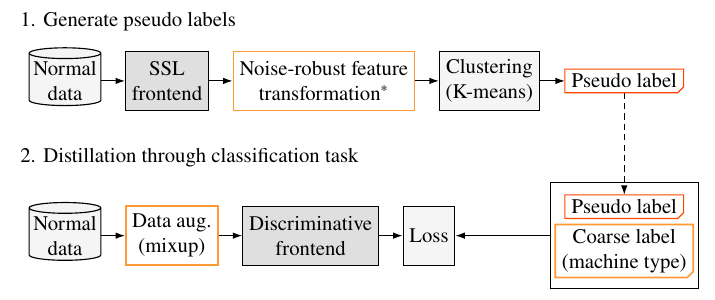}
    \caption{
        Overview of pseudo-label distillation.
        First, \ac{ssl} features are extracted and optionally further transformed into noise-robust features ($*$: requires a small number of clean machine sounds or isolated noise samples).
        Next, pseudo labels are generated by applying clustering to the resulting features.
        Finally, a compact discriminative frontend is trained using the pseudo labels together with available coarse labels and data augmentation.
    }
    \label{fig:overview}
\end{figure}

\Ac{ssl} is a promising label-free approach~\cite{saengthong2025deep,saengthong2026sub,wilkinghoff2026temporal,wilkinghoff2023on}.
\Ac{ssl} models are trained on large-scale unlabeled audio datasets to learn general-purpose audio representations~\cite{chen2023beats,chen2024eat,dinkel2024scaling,fan2026fisher,niizumi2024masked,gong2022ssast}.
These representations effectively capture acoustic characteristics and have achieved strong performance across various audio tasks, including \ac{asd}.
Although discriminative fine-tuning on machine sound datasets using machine-information labels can further improve performance~\cite{jiang2024anopatch,zheng2024improving,han2025exploring}, the original \ac{ssl} features already provide competitive performance without using any label information~\cite{saengthong2025deep,saengthong2026sub,wilkinghoff2026temporal}.
However, \ac{ssl} models are computationally expensive.
Typical \ac{ssl} models~\cite{chen2023beats,chen2024eat,niizumi2024masked,gong2022ssast} have approximately 90M parameters, and recent scaled-up models exceed 1B parameters~\cite{dinkel2024scaling}.
In contrast, typical discriminative feature extractors~\cite{wilkinghoff2023design,wilkinghoff2024self,fujimura2025improvements,liu2022anomalous} have only a few million parameters, corresponding to less than 10\% of the parameter count of a typical \ac{ssl} model.

In this paper, we introduce pseudo-label distillation to address the insufficient performance obtained using coarse labels and the high computational cost of \ac{ssl} models (see \cref{fig:overview}).
This approach first extracts \ac{ssl} features from training data consisting of normal machine sounds and then generates pseudo labels by applying clustering to the extracted features.
Next, a smaller discriminative frontend is trained using these pseudo labels.
During training, available coarse labels (e.g., machine-type labels) and data augmentation are jointly used to further improve performance.
Additionally, we propose lightweight \ac{nrft} methods as a preprocessing step to suppress the effects of noise contained in the training data.
\Ac{nrft} uses a linear transformation estimated using a small amount of clean machine-sound data or isolated noise data.

The pseudo-label distillation approach itself was proposed in our previous work~\cite{fujimura2025improvements}.
Subsequent studies have also explored related techniques, including iterative pseudo-label generation in a discriminative feature space~\cite{kuroyanagi2025improving}, multi-pseudo-label assignment~\cite{chen2026improving}, and domain-adaptive pre-training of teacher \ac{ssl} models~\cite{fang2026improving}.
In contrast, this paper focuses on the original, simple pseudo-label distillation framework and conducts comprehensive evaluations and analyses to provide a deeper understanding.
Our contributions are summarized as follows:
\footnote{Our previous work~\cite{fujimura2025improvements} introduced pseudo-label distillation as one of several improvements to discriminative \ac{asd} methods and investigated it only in a limited experimental setting. In contrast, this work focuses specifically on pseudo-label distillation and provides a more focused background discussion and extensive analyses of this framework. Specifically, the previous work evaluated the method on only two datasets and did not employ \acs{sota} \ac{ssl} models as teacher models. It also did not include an analysis of the effect of the number of clusters used for pseudo-label generation, computational cost analyses, ablation studies on the use of coarse labels and data augmentation, or evaluation of the effectiveness of pseudo-label distillation when fine-tuning the teacher models. In addition, the \acs{nrft} methods are newly proposed in this work.} %
\begin{itemize}
\item We conduct comprehensive evaluations across the DCASE 2020--2025 datasets using four \ac{sota} \ac{ssl} models. The experimental results show that pseudo-label distillation effectively transfers the performance of \ac{ssl} models to smaller discriminative frontends while reducing the number of parameters and \acp{mac} to less than 10\% of those of the original \ac{ssl} models.
\item We show that pseudo-label distillation can achieve even better performance than the original \ac{ssl} models by utilizing available coarse labels and data augmentation.
\item We further show that the performance gains from pseudo-label distillation are observed not only when training small discriminative frontends from scratch but also when fine-tuning \ac{ssl} models.
\item We propose lightweight \ac{nrft} methods that construct noise-robust feature spaces for pseudo-label generation using a small amount of clean machine sounds or isolated noise samples.
\item We publish the code in ASDKit\footnote{\url{https://github.com/TakuyaFujimura/dcase-asd-toolkit}}~\cite{fujimura2025asdkit}.
\end{itemize}

\section{Related work}
Many methods have been proposed to obtain effective feature spaces for \ac{asd}.
Discriminative approaches have been improved in terms of loss functions, data augmentation, and network architectures.
In terms of loss functions, angular-margin losses such as ArcFace~\cite{deng2019arcface} and AdaCos~\cite{zhang2019adacos} have been commonly employed for the \ac{asd} task.
These losses encourage the features of each class to form compact and well-separated clusters, which has been shown to be effective for the downstream \ac{asd} task~\cite{wilkinghoff2023why}.
In addition, less restrictive feature distributions have been shown to further improve \ac{asd} performance.
Such distributions can be learned using \ac{scac}~\cite{wilkinghoff2023why}, which introduces multiple class centers for each class, and trainable class centers rather than fixed centers~\cite{fujimura2025asdkit}.
For data augmentation, mixup~\cite{zhang2018mixup} has been widely used and shown to improve \ac{asd} performance~\cite{fujimura2025asdkit}.
Regarding network architectures, lightweight CNN-based networks have been commonly used~\cite{wilkinghoff2023design,wilkinghoff2024self,liu2022anomalous,kawaguchi2021description,dohi2022description,sandler2018mobilenetv2}.
The DCASE Challenge Task 2 baseline~\cite{kawaguchi2021description,dohi2022description} employed MobileNetV2~\cite{sandler2018mobilenetv2} with log-mel spectrogram input, and an extension using waveform input has been shown to improve performance~\cite{liu2022anomalous}.
Multi-branch architectures comprising spectrogram and spectrum branches have also been widely adopted~\cite{wilkinghoff2023design,wilkinghoff2024self}, with further improvements reported using multi-resolution spectrogram inputs~\cite{fujimura2025improvements}.
These discriminative models are trained from scratch on machine-sound datasets and are considerably more lightweight than \ac{ssl} models.

For \ac{ssl}-based approaches, many \ac{ssl} models have been employed for \ac{asd}.
BEATs~\cite{chen2023beats} is one of the most widely used \ac{ssl} models for various audio tasks, including \ac{asd}~\cite{saengthong2026sub,wilkinghoff2026temporal,jiang2024anopatch,fujimura2025discriminative}.
It is trained through masked audio prediction in a discrete token space, with the tokenizer and \ac{ssl} model iteratively updated through knowledge distillation.
EAT~\cite{chen2024eat} is another \ac{ssl} model widely used for \ac{asd}~\cite{saengthong2026sub,wilkinghoff2026temporal,fujimura2025discriminative}.
It is based on a masked bootstrapping framework and employs a CLS token and an utterance-level loss to capture global information.
Dasheng is a recent large-scale \ac{ssl} model with approximately 1.2B parameters.
It is trained using a masked autoencoder framework on 272k hours of audio data.
These audio \ac{ssl} models are trained on diverse audio datasets spanning speech, music, and environmental sounds and have been shown to achieve better \ac{asd} performance than speech- and music-specific \ac{ssl} models~\cite{han2025exploring,fan2026fisher}.
More recently, \ac{ssl} models specifically designed for machine signals have also been proposed~\cite{zhang2026echo,fan2026fisher}.
These models are designed to handle arbitrary sampling rates and have shown strong performance on fault-diagnosis benchmarks.
Despite these advantages, the aforementioned general-purpose audio \ac{ssl} models have still demonstrated comparable or better \ac{asd} performance~\cite{fan2026fisher}, highlighting their strong generalization capabilities.

\section{Pseudo-label distillation}
\subsection{Overview}
Pseudo-label distillation first extracts \ac{ssl} features from normal training data.
We employ the following four \ac{ssl} models: BEATs~\cite{chen2023beats} (\textit{\path{BEATs_iter3.pt}}), a version of BEATs fine-tuned for audio tagging (BEATs-FT; \textit{\path{BEATs_iter3_plus_AS2M_finetuned_on_AS2M_cpt1.pt}}), EAT~\cite{chen2024eat} (\textit{\path{EAT-base_epoch10_pt.pt}}), and Dasheng~\cite{dinkel2024scaling} (\textit{\path{dasheng-1.2B}}). %
These \ac{ssl} models receive a sequence of mel-spectrogram patches or segments and output the corresponding feature sequence.
For BEATs and Dasheng, we average the \ac{ssl} feature sequence to obtain a single feature vector for each audio sample, whereas for EAT, we use the CLS token.
To generate pseudo labels, we apply standard k-means clustering, with the number of clusters determined from the number of training samples using the cluster ratio $r$. %
We evaluate several values of $r$, as well as, for reference, an oracle setting in which the best value of $r$ is selected separately for each machine type. %

To not only reduce computational cost but also further improve performance, we jointly use available coarse labels during pseudo-label distillation.
In the experiments, we use the machine-type labels provided in the DCASE datasets~\cite{koizumi2020description,kawaguchi2021description,dohi2022description,dohi2023description,nishida2024description,nishida2025description} as coarse labels.
Using these coarse labels, we perform clustering separately for each coarse-label class and combine the resulting pseudo labels to construct a multi-class classification task for distillation.
For example, the DCASE 2023 dataset contains 14 machine types, each with 1,000 normal training samples.
Thus, setting $r=0.2\%$ yields 2 pseudo-label classes per machine type, resulting in a multi-class classification task with 28 classes in total.
Additionally, motivated by the Noisy Student framework~\cite{xie2020self}, we employ mixup as data augmentation during pseudo-label distillation.

\subsection{Noise-robust feature transformation (NRFT)}
\label{sec:noise_robust_feature_transformation}
In the experimental evaluations in~\cref{sec:experiments}, we show that pseudo-label distillation can sometimes degrade performance.
This degradation occurs because the training data consist of noisy machine sounds, and the \ac{ssl} features capture not only machine characteristics but also noise characteristics.
When noise conditions vary across training samples and the noise dominates the machine sound, the resulting pseudo labels primarily reflect differences in noise conditions rather than machine characteristics.
Such noise-dependent pseudo labels make the student frontend sensitive to noise variations, thereby degrading performance.

To address this issue, we propose a lightweight \acf{nrft} method for pseudo-label generation.
In general, improving noise robustness requires auxiliary information about either noise or clean target sounds.
In this work, we consider the setting of DCASE 2025 Task~2~\cite{nishida2025description}, where a small number of either clean machine-sound samples or isolated noise samples are available for each machine type.

For the case where clean machine sounds are available, we propose a \ac{pca}-based subspace projection.
We first extract \ac{ssl} features from the clean machine sounds and apply \ac{pca} to the extracted clean features.
We then construct a projection matrix $\bm{V}\in\mathbb{R}^{D\times K}$ using the top $K$ principal components, where $D$ denotes the feature dimension.
The \ac{ssl} features of the noisy machine sounds in the training data are projected onto this subspace, and clustering is then performed to obtain pseudo labels.
This projection suppresses components that are less representative of the clean machine sounds and are therefore likely to reflect noise-related variations.

For the case where isolated noise samples are available, we propose a \ac{gevd}-based method.
We aim to obtain a feature transformation that preserves variations in noisy machine sounds while suppressing directions dominated by noise.
The corresponding transformation direction $\bm{v}\in\mathbb{R}^{D}$ is obtained by solving the following optimization problem:
\begin{equation}
\label{eq:noise_robust_subspace}
\max_{\bm{v}} \frac{\bm{v}^\top \bm{R}_{\mathrm{x}} \bm{v}}
{\bm{v}^\top \bm{R}_{\mathrm{n}} \bm{v}},
\end{equation}
where $\bm{R}_{\mathrm{x}}\in\mathbb{R}^{D\times D}$ and $\bm{R}_{\mathrm{n}}\in\mathbb{R}^{D\times D}$ denote the covariance matrices of the \ac{ssl} features of the noisy machine sounds and the isolated noise samples, respectively.
This problem is solved by \ac{gevd}, leading to the following problem:
\begin{equation}
\bm{R}_{\mathrm{x}} \bm{v} = \lambda \bm{R}_{\mathrm{n}} \bm{v}.
\end{equation}
The eigenvalue $\lambda$ represents the ratio of the variance of the noisy machine sound features to that of the noise features along direction $\bm{v}$.
The transformation matrix $\bm{V}\in\mathbb{R}^{D\times K}$ is constructed from the eigenvectors corresponding to the $K$ largest eigenvalues.
These eigenvectors are normalized such that $\bm{V}^{\top}\bm{R}_{\mathrm{n}}\bm{V}=\bm{I}$, where $\bm{I}$ denotes the identity matrix.
When $K<D$, this transformation projects the features onto a noise-robust subspace.
When $K=D$, the transformation corresponds to whitening with respect to the noise covariance matrix $\bm{R}_{\mathrm{n}}$, without dimensionality reduction.
In practice, due to the limited size of the isolated-noise dataset, we first perform \ac{pca} using the features of both the noisy machine sounds and isolated noise samples, reducing the feature dimension to $D=64$.
We then solve \cref{eq:noise_robust_subspace} in the reduced feature space.

\section{Experimental evaluation}
\label{sec:experiments}
We conduct extensive analyses of pseudo-label distillation.
In~\cref{sec:results_comprehensive}, we comprehensively evaluate pseudo-label distillation across the DCASE 2020--2025 datasets using four \ac{ssl} models and analyze its computational cost.
In~\cref{sec:results_ablation}, we analyze the effects of using coarse machine-type labels and mixup during pseudo-label distillation.
In~\cref{sec:results_finetuning}, we also evaluate the use of pseudo labels for fine-tuning \ac{ssl} models, focusing on performance improvement rather than computational cost reduction.
In~\cref{sec:results_noise_robust}, we evaluate our \ac{nrft} methods.

\subsection{Setups}
\label{sec:setup}
We conducted experimental evaluations on the DCASE 2020--2025 Task~2 datasets~\cite{koizumi2020description,kawaguchi2021description,dohi2022description,dohi2023description,nishida2024description,nishida2025description}.
These datasets are constructed from different data sources: 
\cite{purohit2019mimii} and \cite{koizumi2019toyadmos} for 2020;
\cite{tanabe2021mimii} and \cite{harada2021toyadmos2} for 2021;
\cite{dohi2022mimii} and \cite{harada2021toyadmos2} for 2022;
\cite{dohi2022mimii}, \cite{harada2021toyadmos2}, and \cite{harada2023toyadmos2} for 2023;
\cite{dohi2022mimii}, \cite{harada2021toyadmos2}, \cite{niizumi2024toyadmos2}, and \cite{albertini2024imad} for 2024;
and \cite{dohi2022mimii}, \cite{harada2021toyadmos2}, and \cite{harada2025toyadmos2025} for 2025.
The recordings are 6--18-second single-channel audio samples recorded at 16~kHz.
Each dataset contains several machine types.
The DCASE 2020--2022 datasets include 6--7 machine types, and each machine type has 6--7 subsets called \textit{sections}.
Each section consists of training data with approximately 1,000 normal-sound samples and test data with approximately 100--200 samples for each of the normal and anomalous conditions.
The DCASE 2023--2025 datasets include 14--16 machine types, and each machine type has a single section.
The DCASE 2025 dataset additionally provides 100 samples of clean machine sounds or isolated noise for each machine type, which can be used for \ac{nrft}.
For all datasets, sections and machine types are grouped into \textit{dev} and \textit{eval} subsets, and the evaluation results are aggregated and reported separately for each subset.

We followed the official evaluation metrics for each dataset.
In DCASE 2020, \ac{auc} and \ac{pauc} with $p=0.1$ are calculated for each section.
These scores are then aggregated by taking the arithmetic mean across sections and machine types to obtain machine-type-wise scores and dev/eval-wise scores.
For DCASE 2021--2025, a domain shift problem~\cite{wilkinghoff2025handling} has been introduced, in which the recording environments and machine operating conditions differ between the \textit{source} and \textit{target} domains.
While the \textit{target} domain includes only approximately 1\% as many training samples as the \textit{source} domain, the test data include the same number of samples from both domains, requiring models to generalize across domains.
In DCASE 2021, \ac{auc} and \ac{pauc} are calculated for each domain.
These scores are then aggregated by taking the harmonic mean across sections, domains, and machine types.
In DCASE 2022--2025, \ac{auc} is calculated for each domain, whereas \ac{pauc} is calculated using samples from both domains.
These scores are aggregated using the harmonic mean.
Further details of the task setups are provided in the original papers~\cite{koizumi2020description,kawaguchi2021description,dohi2022description,dohi2023description,nishida2024description,nishida2025description}, and the setups for DCASE 2020--2024 are also summarized in~\cite{fujimura2025asdkit}.
We additionally report average evaluation scores across the different datasets for reference. Note that these averages are based on scores obtained under different evaluation protocols and aggregation rules.

\begin{table}[t]
    \centering
    \caption{Methods used in the experiments. The suffix of each \textit{Dis-single-*}, \textit{Dis-multi-*}, and \textit{Dis-SSL-*} is either \textit{machine}, \textit{GT}, \textit{PL}, or \textit{NRFT}, indicating the use of coarse machine-type labels, detailed ground-truth labels, pseudo labels, or pseudo labels generated with \ac{nrft}, respectively.}
    \label{tab:methods}
    \resizebox{\columnwidth}{!}{
    \begin{tabular}{ll}
    \toprule
    Method name & Description \\
    \midrule
    Dis-single-* & Discriminative frontend using a single-resolution spectrogram \\
    Dis-multi-* & Discriminative frontend using multi-resolution spectrograms \\
    Raw-SSL & Direct use of original \ac{ssl} features \\
    Dis-SSL-* & Discriminative fine-tuning of an \ac{ssl} model \\
    \bottomrule
    \end{tabular}
    }
\end{table}

\begin{figure*}[tb]
\begin{center}
\includegraphics[width=\linewidth]{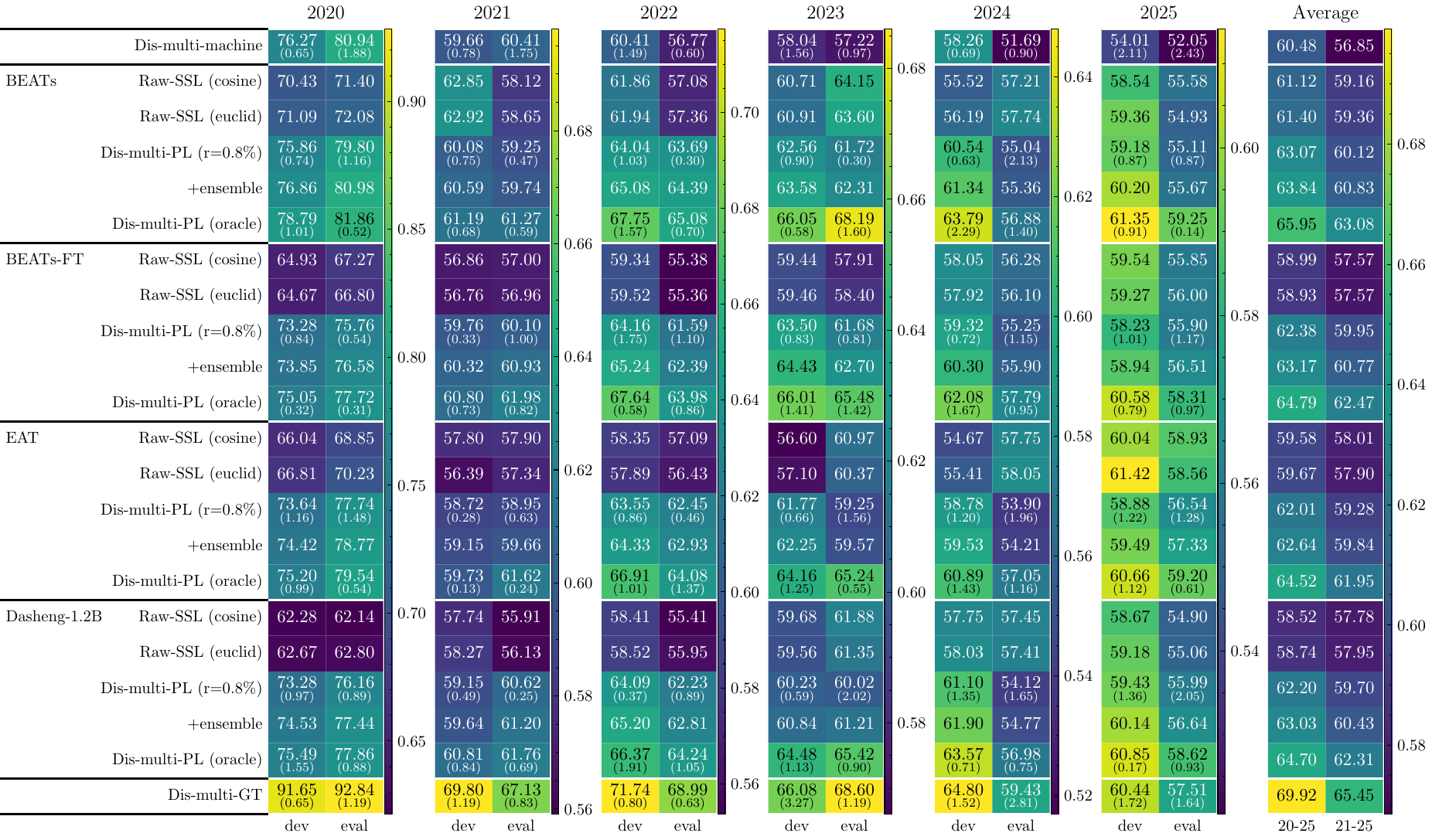}
\end{center}
\caption{
Official evaluation scores on the DCASE 2020--2025 datasets using four \ac{ssl} models.
The scores are reported as ``average (CI95)'' across four trials.
Additionally, average scores across all datasets (DCASE 2020--2025) and across DCASE 2021--2025 are reported separately because DCASE 2020 exhibits a different trend from those of the other datasets.
}
\label{fig:official_score}
\end{figure*}

\begin{figure}[tb]
\begin{center}
\includegraphics[width=0.95\columnwidth]{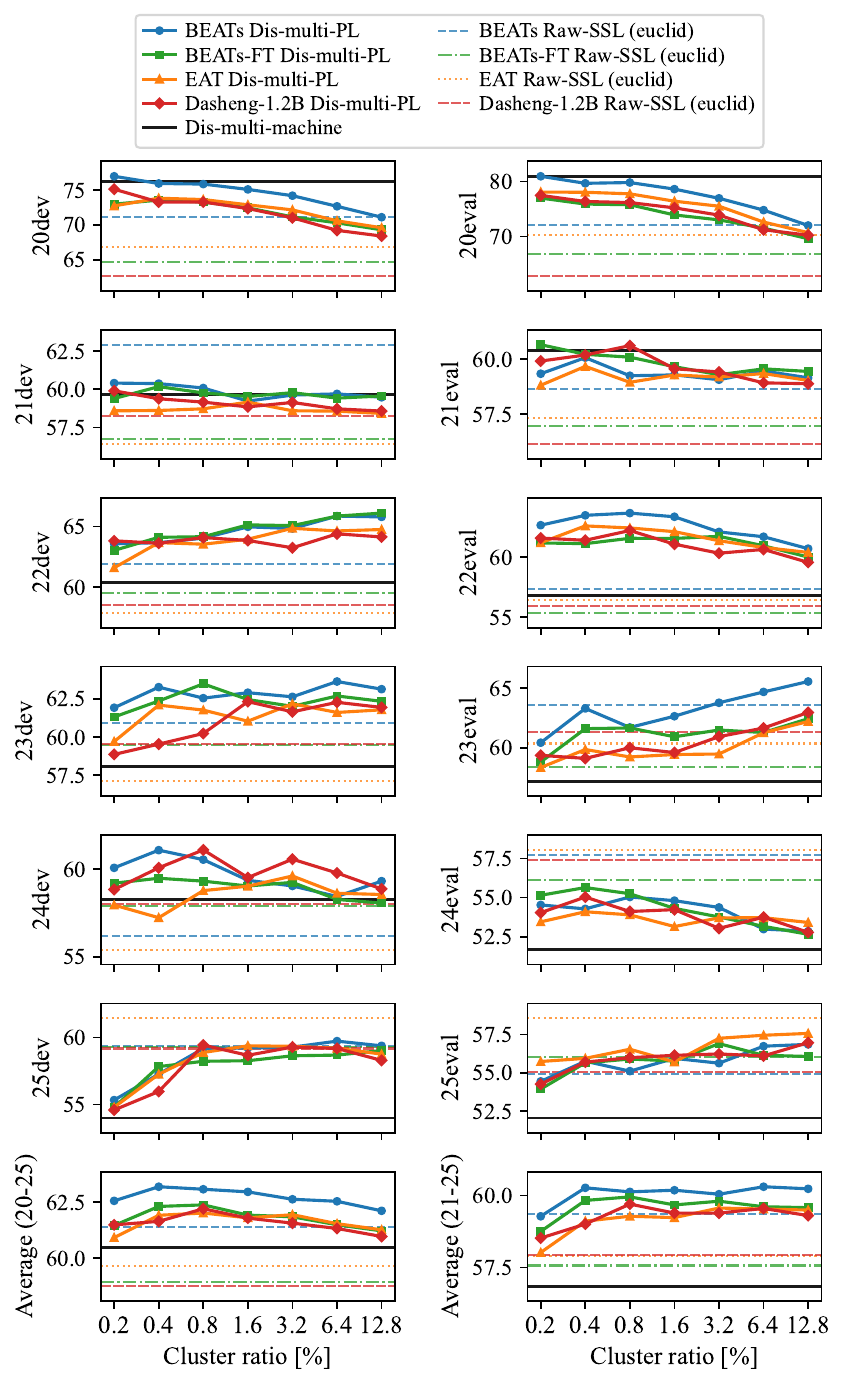}
\end{center}
\caption{
Evaluation results for different cluster ratios $r$.
The average official evaluation scores across four trials are shown for each cluster ratio.
Additionally, average scores across all datasets (DCASE 2020--2025) and across DCASE 2021--2025 are reported separately because DCASE 2020 exhibits a different trend from those of the other datasets.
}
\label{fig:ssl_official}
\end{figure}

The DCASE datasets include coarse machine-type labels and more detailed labels (e.g., machine IDs and attributes).
To simulate unlabeled conditions, we concealed all labels except for the machine-type labels.
In the DCASE 2024 and 2025 datasets, approximately half of the machine types are originally provided with only machine-type labels, and we also conducted evaluations using these public benchmark setups.
For pseudo-label generation, we varied the cluster ratio $r$ over 0.2, 0.4, 0.8, 1.6, 3.2, 6.4, and 12.8\%.

For the student discriminative frontend, we used two multi-branch architectures~\cite{wilkinghoff2023design,wilkinghoff2024self,fujimura2025improvements}.
One used spectrum and single-resolution spectrogram inputs computed using a \ac{stft} with a \ac{dft} size of 1024, whereas the other used spectrum and multi-resolution spectrogram inputs~\cite{fujimura2025improvements} computed using \ac{stft}s with \ac{dft} sizes of 256, 1024, and 4096.
The hop size for each \ac{stft} was set to half the corresponding \ac{dft} size.
We trained each network for 16 epochs using \ac{scac} with 16 sub-clusters, the AdamW optimizer~\cite{loshchilov2019decoupled}, a fixed learning rate of 0.001, and a batch size of 64.
The mixup probability was set to 0.5.

For fine-tuning BEATs and EAT in \cref{sec:results_finetuning}, we used \ac{lora}~\cite{hu2022lora}.
For BEATs, we introduced \ac{lora} parameters into the query and key projections in the Transformer layers.
The output feature sequence was transformed into a single 256-dimensional discriminative feature using an attentive statistics pooling layer~\cite{okabe2018interspeech} and a linear layer.
For EAT, we introduced \ac{lora} parameters into the query, key, and value projections.
The CLS token was transformed into a single 256-dimensional discriminative feature using a linear layer.
We fine-tuned the models for 25 epochs using \ac{scac}, the AdamW optimizer, a batch size of 8, and a \ac{lora} rank of 64.
During the initial 5,000 steps, the learning rate was increased linearly from 0 to 0.0001.

For the backend, we employed a standard nearest-neighbor-based approach with SMOTE oversampling~\cite{fujimura2025asdkit}.
We calculated the nearest-neighbor distance between a test sample and normal training samples in the feature space as the anomaly score.
To mitigate the data imbalance between the source and target domains, SMOTE oversampling~\cite{chawla2002smote} was applied to the target-domain training samples in the feature space.
Specifically, we oversampled the target-domain samples using two nearest neighbors until the number of target-domain samples reached 20\% of the number of source-domain samples.

All discriminative training was performed across four trials with different neural-network initializations. %
We confirmed that the variation caused by different k-means initializations was comparable to that caused by different neural-network initializations and used the same set of pseudo labels generated with the same k-means initialization for all trials. %
We report the average performance and the 95\% confidence interval (CI95) across the trials. %
Also, we show the ensemble results obtained by averaging the anomaly scores across the trials.

\Cref{tab:methods} summarizes the methods and their names used in the experiments.
\textit{Dis-single-*} and \textit{Dis-multi-*} denote discriminative frontends using single- and multi-resolution spectrograms, respectively, where the suffix indicates the labels used for training: \textit{machine} for coarse machine-type labels, \textit{GT} for detailed ground-truth labels, \textit{PL} for pseudo labels, and \textit{NRFT} for pseudo labels generated with \ac{nrft}.
\textit{Raw-SSL} denotes the direct use of the original \ac{ssl} features, and \textit{Dis-SSL-*} denotes discriminative fine-tuning of an \ac{ssl} model with the corresponding labels.
Since we used \ac{scac}, a cosine-distance-based classification loss, for discriminative training, cosine distance was used in the backend for Dis-single, Dis-multi, and Dis-SSL.
For Raw-SSL, we evaluated both cosine and Euclidean distances.

\subsection{Results}
\label{sec:results}
\subsubsection{Comprehensive evaluation}
\label{sec:results_comprehensive}
\Cref{fig:official_score} shows the evaluation results on the DCASE 2020--2025 datasets, where Dis-multi-* was used as the student discriminative frontend.
First, the large performance gap between the models trained with machine labels and those trained with GT labels indicates that coarse machine labels alone are insufficient for learning a feature space effective for \ac{asd}.
On the other hand, Raw-SSL methods achieve competitive performance without using any label information.
For example, on DCASE 2025, all Raw-SSL methods outperform Dis-multi-machine.
Moreover, Raw-SSL with BEATs outperforms Dis-multi-machine in terms of average performance across all datasets.
These results support our motivation for introducing pseudo-label distillation.

Pseudo-label distillation effectively transfers the performance of Raw-SSL to the discriminative frontends.
In \cref{fig:official_score}, the results with a fixed cluster ratio of $r=0.8\%$ are presented as a representative setting for Dis-multi-PL. %
Dis-multi-PL improves performance over Dis-multi-machine, especially on the DCASE 2022--2025 datasets, with the same inference-time computational cost as Dis-multi-machine, demonstrating the benefit of distilling Raw-SSL. %
During inference, the SSL teacher is no longer required, and only the compact student frontend is used. %
Notably, Dis-multi-PL not only transfers the performance of Raw-SSL but also further improves it. %
On the DCASE 2022 eval subset, Dis-multi-machine achieves 56.77\%, Raw-SSL with BEATs achieves 57.08\%, and Dis-multi-PL achieves 63.69\%, yielding an improvement of approximately 6 percentage points. %
Also, Dis-multi-PL achieves the best average performance across all datasets for each of the \ac{ssl} models.
Even when excluding the exceptional case of DCASE 2020, in which machine labels provide a large performance gain and Dis-multi-machine and Dis-multi-PL substantially outperform Raw-SSL, Dis-multi-PL still achieves the best average performance, demonstrating its general effectiveness.
We analyze the factors behind this performance improvement in \cref{sec:results_ablation}.
The performance of pseudo-label distillation is further improved by ensembling the results of multiple training trials, which are easier to perform owing to the low computational cost of the student discriminative frontends.
Also, the results obtained using the oracle cluster ratio demonstrate the potential for substantial performance improvement by selecting an appropriate cluster ratio for each machine type.

\Cref{fig:ssl_official} shows the sensitivity of pseudo-label distillation to the cluster ratio $r$.
Although the optimal cluster ratio varies across datasets, performance improvements of Dis-multi-PL over Dis-multi-machine can be observed across a wide range of cluster ratios, especially on the DCASE 2022--2025 datasets.
Also, on the DCASE 2022 dataset, Dis-multi-PL outperforms Raw-SSL across multiple values of $r$.
The average results across multiple datasets show a similar trend.
Although Dis-multi-PL is equivalent to Dis-multi-machine when the resulting number of classes for each machine type is one, performance improvements are observed even with small cluster ratios such as $r=0.2\%$ and $0.4\%$.
For example, the DCASE 2023--2025 datasets contain 1,000 training samples for each machine type, and $r=0.2\%$ results in only two classes.
These results indicate that the effectiveness of pseudo-label distillation is not restricted to a narrowly tuned value of $r$, although the optimal value varies across datasets.
In the following experiments, we use $r=0.8\%$, which is of a similar order to the ground-truth class ratios in the DCASE datasets, as a representative fixed setting. %

\begin{table}[tb]
    \centering
    \caption{Computational cost, measured in \acp{mac}, for a 10-second input, and the number of parameters.}
    \label{tab:model_size}
    \resizebox{\columnwidth}{!}{
    \begin{tabular}{lrrrrr}
    \toprule
    & BEATs & EAT & Dasheng & Dis-single & Dis-multi\\
    \midrule
    \acs{mac}s (G) & 44.81 & 43.71 & 283.34 & 0.44 & 1.17 \\
    Params (M) & 90.31 & 85.25 & 1133.66 & 2.50 & 5.00 \\
    \bottomrule
    \end{tabular}
    }
\end{table}

\Cref{tab:model_size} shows the number of \acp{mac} required to process a 10-second audio signal and the number of parameters for each frontend.
BEATs and EAT require over 40~G\acp{mac} and 85M parameters, and Dasheng requires even more computational resources.
In contrast, Dis-multi requires only 1.17~G\acp{mac} and 5M parameters.
Moreover, Dis-single further reduces the computational cost; we show that it still achieves competitive performance in \cref{sec:results_ablation}.

\begin{table}[tb]
    \centering
    \caption{Evaluation results on the original DCASE 2024 and 2025 datasets.
    ORG denotes the original label setting with partially missing labels.
    The official evaluation scores are reported as ``average (CI95)'' across four trials.
    }
    \label{tab:evaluation_2425}
    \resizebox{\columnwidth}{!}{
    \begin{tabular}{lllllll}
        \toprule
        \multicolumn{2}{l}{Method} & 24dev & 24eval & 25dev & 25eval & Average \\
        \midrule
        \multicolumn{2}{l}{Dis-multi-ORG} & \makecell{63.05 \\[-3pt] {\scriptsize (1.58)}} & \makecell{56.73 \\[-3pt] {\scriptsize (2.93)}} & \makecell{58.97 \\[-3pt] {\scriptsize (0.80)}} & \makecell{54.86 \\[-3pt] {\scriptsize (4.15)}} & 58.40 \\
        \midrule
        BEATs & Raw-SSL (euclid) & 56.19 & 57.74 & 59.36 & 54.93 & 57.06 \\
        & Dis-multi-PL ($r=0.8\%$) & \makecell{63.28 \\[-3pt] {\scriptsize (1.31)}} & \makecell{55.97 \\[-3pt] {\scriptsize (1.07)}} & \makecell{60.78 \\[-3pt] {\scriptsize (1.50)}} & \makecell{57.69 \\[-3pt] {\scriptsize (0.89)}} & 59.43 \\
        & Dis-multi-PL (oracle) & \makecell{\textbf{65.49} \\[-3pt] {\scriptsize (1.96)}} & \makecell{\textbf{60.17} \\[-3pt] {\scriptsize (1.45)}} & \makecell{62.63 \\[-3pt] {\scriptsize (0.93)}} & \makecell{\textbf{60.13} \\[-3pt] {\scriptsize (1.17)}} & \textbf{62.11} \\
        \midrule
        EAT & Raw-SSL (euclid) & 55.41 & 58.05 & 61.42 & 58.56 & 58.36 \\
        & Dis-multi-PL ($r=0.8\%$) & \makecell{62.38 \\[-3pt] {\scriptsize (1.18)}} & \makecell{56.26 \\[-3pt] {\scriptsize (1.78)}} & \makecell{59.21 \\[-3pt] {\scriptsize (1.59)}} & \makecell{57.89 \\[-3pt] {\scriptsize (1.23)}} & 58.94 \\
        & Dis-multi-PL (oracle) & \makecell{64.72 \\[-3pt] {\scriptsize (1.76)}} & \makecell{59.46 \\[-3pt] {\scriptsize (2.24)}} & \makecell{\textbf{62.75} \\[-3pt] {\scriptsize (1.27)}} & \makecell{59.77 \\[-3pt] {\scriptsize (1.77)}} & 61.67 \\
        \midrule
        \multicolumn{2}{l}{Dis-multi-GT} & \makecell{64.80 \\[-3pt] {\scriptsize (1.52)}} & \makecell{59.43 \\[-3pt] {\scriptsize (2.81)}} & \makecell{60.44 \\[-3pt] {\scriptsize (1.72)}} & \makecell{57.51 \\[-3pt] {\scriptsize (1.64)}} & 60.54 \\
        \bottomrule
        \end{tabular}
    }
\end{table}

\begin{figure}[tb]
\begin{center}
\includegraphics[width=0.95\columnwidth]{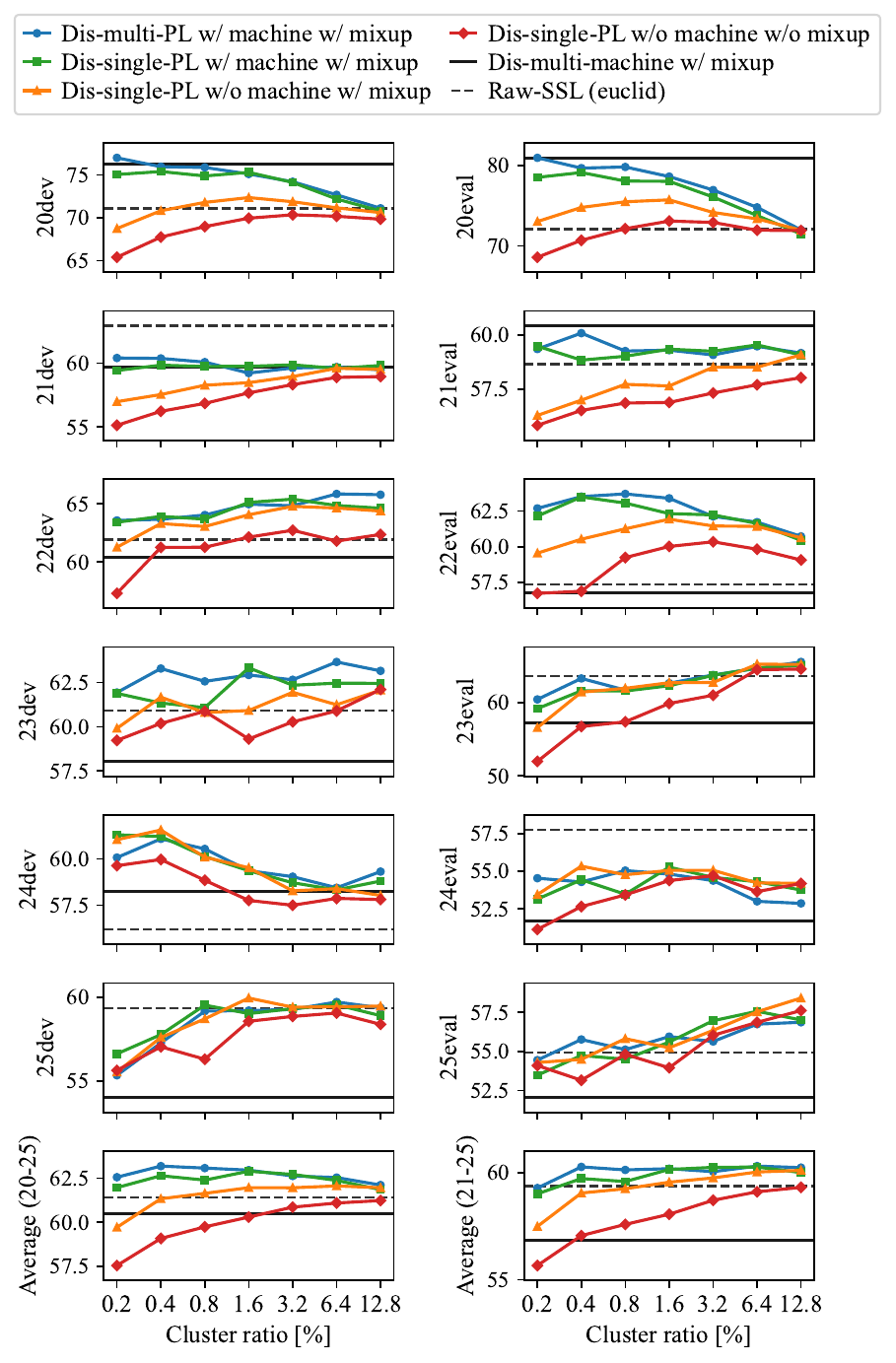}
\end{center}
\caption{
Ablation study on the effects of machine-type labels and mixup in pseudo-label distillation using BEATs.
The average official evaluation scores across four trials are shown for each cluster ratio.
Additionally, average scores across all datasets (DCASE 2020--2025) and across DCASE 2021--2025 are reported separately because DCASE 2020 exhibits a different trend from those of the other datasets.
}
\label{fig:ablation_official}
\end{figure}

\begin{figure}[tb]
\begin{center}
\includegraphics[width=0.95\linewidth]{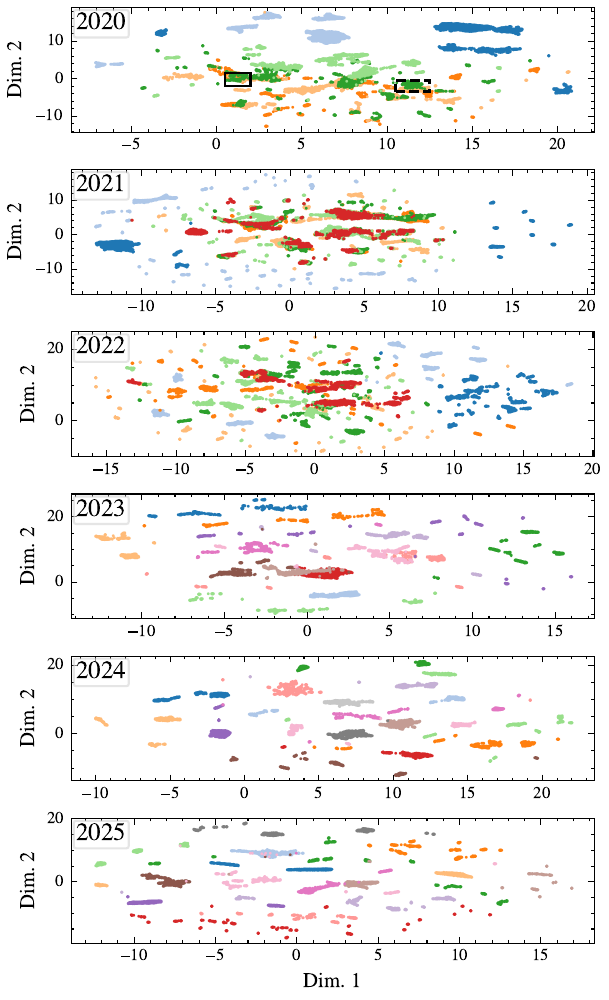}
\end{center}
\caption{
UMAP visualizations of the original BEATs features for the DCASE 2020--2025 datasets.
Colors indicate machine types.
Spectrograms corresponding to the samples from the solid and dashed rectangles in the DCASE 2020 visualization are shown in \cref{fig:spectrograms}.
}
\label{fig:umap}
\end{figure}

\begin{figure}[t!]
\begin{center}
\includegraphics[width=\linewidth]{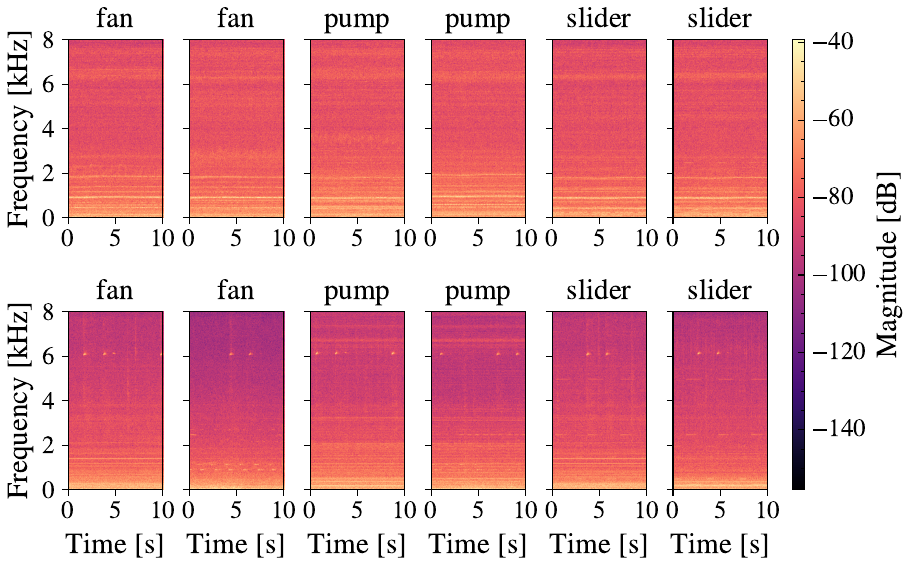}
\end{center}
\caption{
Spectrograms of the samples selected from the solid (top) and dashed (bottom) rectangles in \cref{fig:umap}, respectively.
The spectrograms in the top row exhibit similar stationary noise components, for example, around 1 kHz, whereas those in the bottom row exhibit similar non-stationary noise components around 6 kHz.
}

\label{fig:spectrograms}
\end{figure}

\begin{figure*}[tb]
\begin{center}
\includegraphics[width=\linewidth]{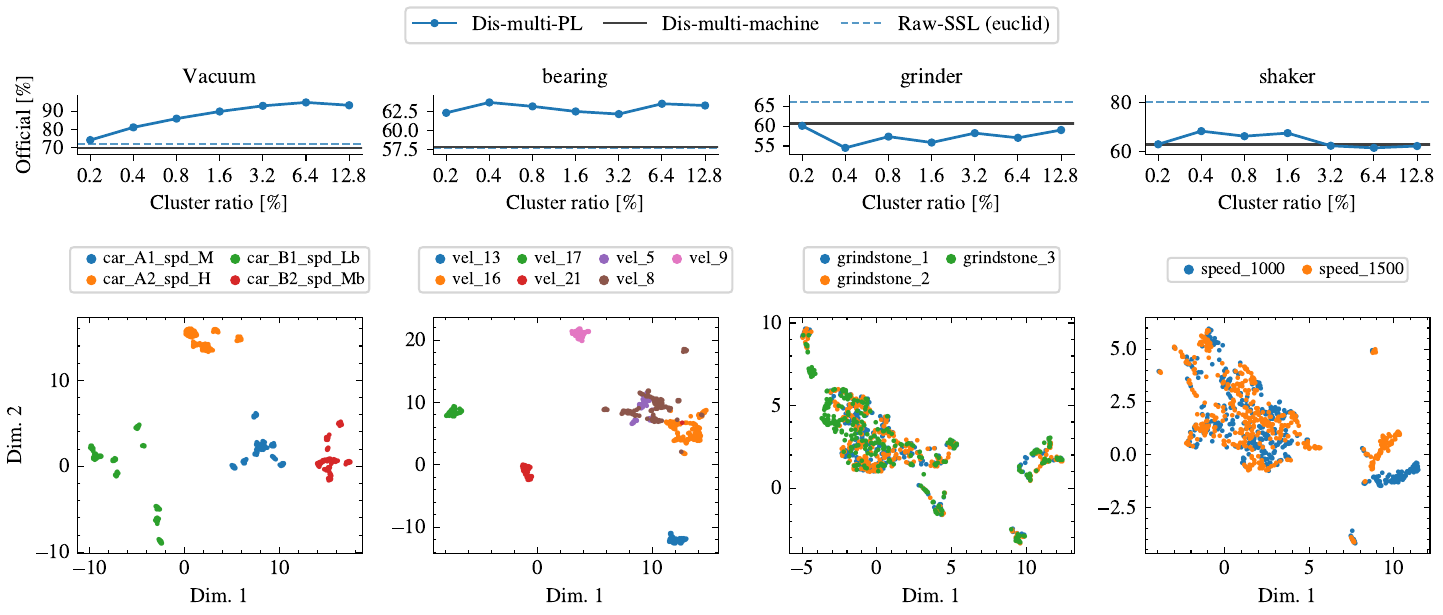}
\end{center}
\caption{
Machine-wise analysis of pseudo-label distillation using BEATs on the DCASE 2023 dataset.
From left to right, the columns correspond to Vacuum, bearing, grinder, and shaker.
The top row shows the official evaluation scores averaged over four trials for each cluster ratio.
The bottom row visualizes the original BEATs feature space of the source-domain training samples, with colors indicating the ground-truth attribute labels.
}
\label{fig:unified_machine_wise_umap_graph}
\end{figure*}

\Cref{tab:evaluation_2425} shows the evaluation results under the original DCASE 2024 and 2025 setups.
We generated pseudo labels for machine types with only machine-type labels, while using the original detailed labels for the other machine types.
From the table, we can still see the effectiveness of pseudo-label distillation, even though detailed labels are available for half of the machine types.
For example, in terms of average performance, Dis-multi-ORG, which uses the original partially missing labels, achieves 58.40\%, Raw-SSL with BEATs achieves 57.06\%, and Dis-multi-PL achieves 59.43\%.
Also, for DCASE 2025, Dis-multi-PL achieves performance comparable to that of Dis-multi-GT by using pseudo labels together with the available detailed labels.
Furthermore, as in \cref{fig:official_score}, pseudo-label distillation substantially improves performance with the oracle cluster ratio.

Overall, the experimental results across multiple settings demonstrate that pseudo-label distillation effectively transfers the benefits of computationally expensive \ac{ssl} models to compact discriminative frontends, while enabling these frontends to outperform Dis-multi-machine and Raw-SSL.

\begin{figure*}[tb]
\begin{center}
\includegraphics[width=\linewidth]{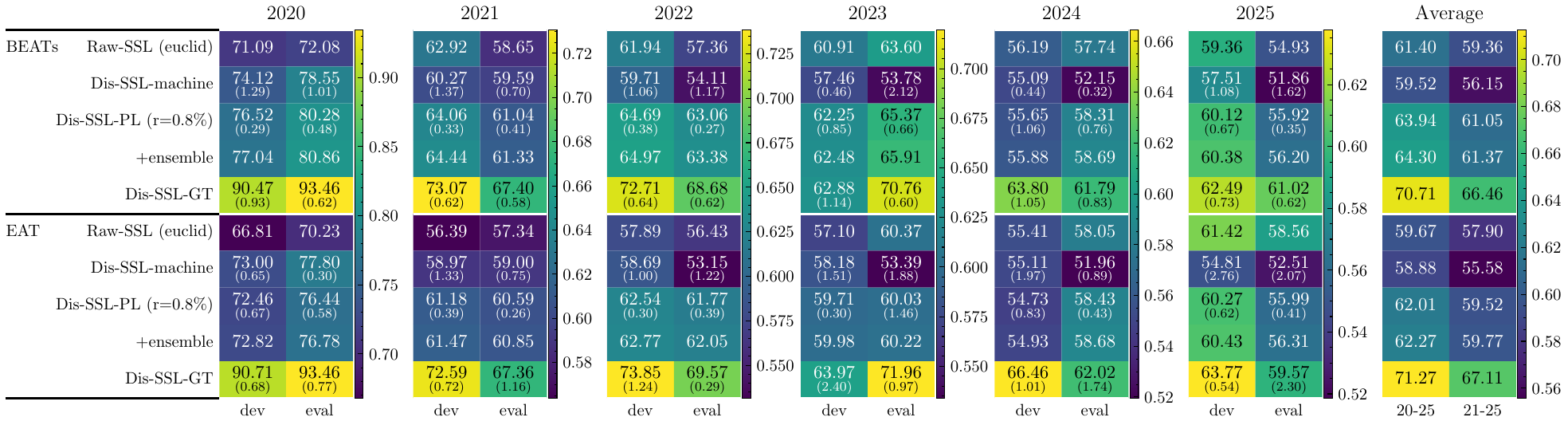}
\end{center}
\caption{
Official evaluation scores of \ac{ssl} models fine-tuned using pseudo labels.
Average scores are reported separately for all datasets (DCASE 2020--2025) and for DCASE 2021--2025 because DCASE 2020 exhibits a different trend from those of the other datasets.
}
\label{fig:disssl_official}
\end{figure*}

\begin{figure*}[t!]

\centering
\includegraphics[width=\linewidth]{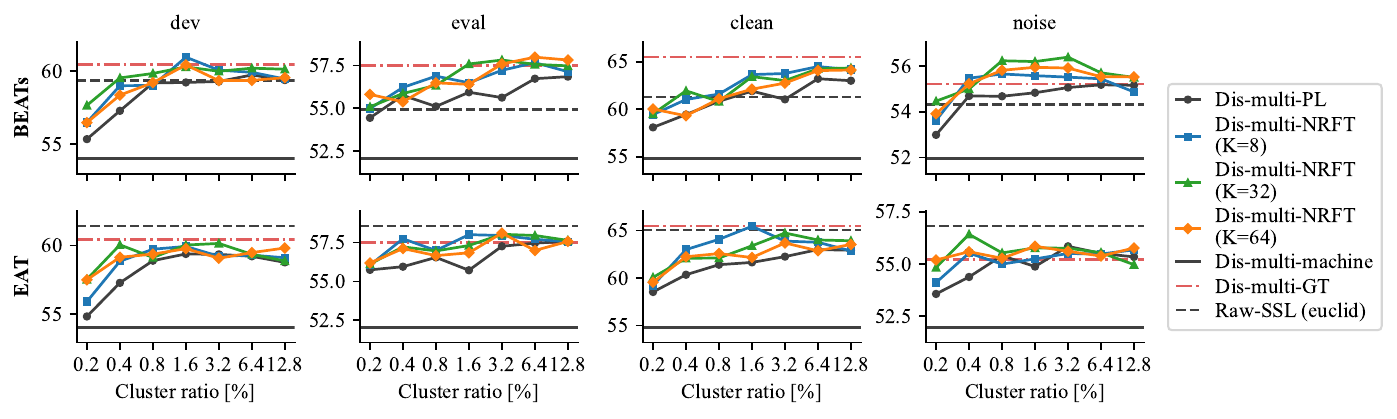}
\caption{
Evaluation results of the \ac{nrft} methods using BEATs and EAT on the DCASE 2025 dataset.
The first and second columns show the results aggregated over the dev and eval subsets, respectively, whereas the third and fourth columns show the results aggregated over the clean-data-available and noise-data-available cases, respectively.
}
\label{fig:dcase2025_hmean}
\end{figure*}

\subsubsection{Analysis of performance gains from pseudo-label distillation}
\label{sec:results_ablation}
\Cref{fig:ablation_official} compares different distillation configurations using BEATs, examining the effects of multi-resolution spectrograms, machine labels, and mixup.
First, although Dis-multi achieves the best performance in most cases, Dis-single also achieves competitive performance while providing a more compact and efficient frontend, as shown in \cref{tab:model_size}.
Second, both machine labels and mixup improve performance, contributing to the student frontend outperforming its teacher SSL model (i.e., Raw-SSL). Also, we can see that these improvements are more pronounced on the DCASE 2020--2022 datasets.
To investigate why their benefits differ across datasets, we analyze the original BEATs feature spaces.
\Cref{fig:umap} shows the UMAP~\cite{mcinnes2018umap} visualizations for the DCASE 2020--2025 datasets.
The visualizations show that, in the DCASE 2020--2022 datasets, the BEATs features of different machine types are more extensively intermixed than those in the other datasets.
\Cref{fig:spectrograms} shows spectrograms of audio samples selected from the solid and dashed rectangles in the DCASE 2020 visualization of \cref{fig:umap}.
These examples indicate that samples from different machine types share similar background noise components and consequently form clusters based on noise characteristics.
In such cases, machine-type classification can encourage the frontend to ignore noise and learn a noise-robust feature space, thereby improving \ac{asd} performance~\cite{wilkinghoff2023why}.
However, this benefit is smaller for DCASE 2023--2025, in which the feature spaces are already separated by machine type.

We also analyze differences in the effectiveness of pseudo-label distillation across machine types.
\Cref{fig:unified_machine_wise_umap_graph} shows machine-wise analyses for four machine types from the DCASE 2023 dataset.
We can see that the characteristics of the feature space differ across machine types, resulting in differences in the effectiveness of pseudo-label distillation and the optimal cluster ratio.
For Vacuum and bearing, the BEATs feature spaces clearly reflect the ground-truth attributes, and pseudo-label distillation substantially improves performance.
In contrast, for grinder and shaker, the feature spaces do not clearly reflect the attributes, resulting in only limited performance gains.

In this section, we showed that machine-type labels and mixup contribute to performance improvements.
We also mentioned that the performance benefits of pseudo-label distillation depend on the \ac{ssl} feature space, particularly on how well it captures the ground-truth attributes and how robust it is to noise variations.

\begin{figure*}[tb]
\begin{center}
\includegraphics[width=0.98\linewidth]{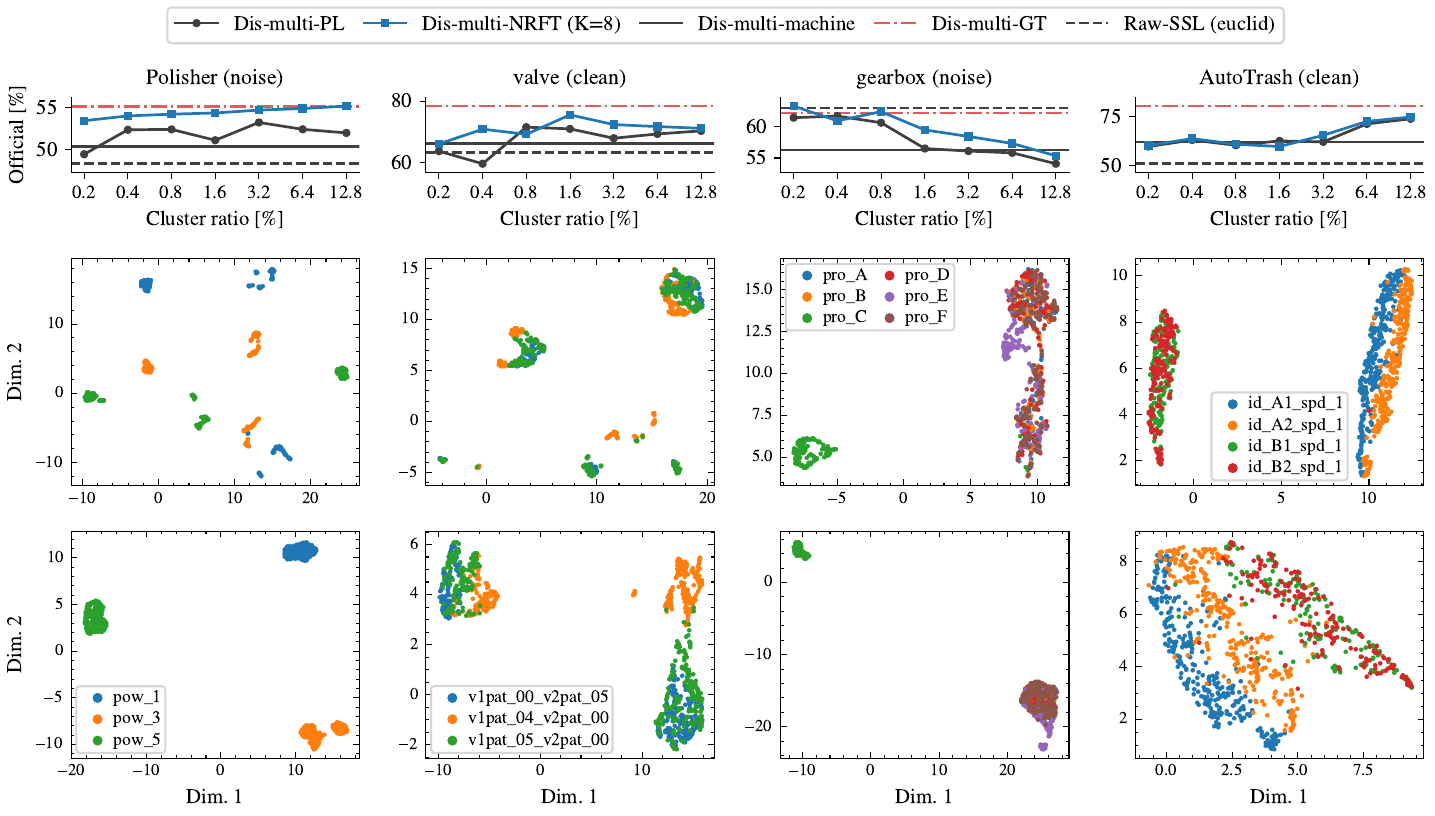}
\end{center}
\caption{
Machine-wise analysis of pseudo-label distillation using BEATs with \ac{nrft} on the DCASE 2025 dataset.
From left to right, the columns correspond to Polisher, valve, gearbox, and AutoTrash.
The labels (clean) and (noise) indicate that clean machine sounds and isolated noise samples, respectively, are available as supplementary data.
The top row shows the official evaluation scores averaged over four trials for each cluster ratio.
The middle and bottom rows visualize the original BEATs feature space and the feature space after \ac{nrft}, respectively, for source-domain training samples, with colors indicating ground-truth attribute labels.
}
\label{fig:unified_machine_wise_umap_graph_dcase2025}
\end{figure*}

\subsubsection{Effectiveness on fine-tuning \ac{ssl} models}
\label{sec:results_finetuning}
\Cref{fig:disssl_official} shows the evaluation results for fine-tuning \ac{ssl} models using pseudo labels, where each \ac{ssl} model is fine-tuned using pseudo labels generated from the original \ac{ssl} model.
First, fine-tuning using only machine labels can degrade performance compared with the Raw-SSL method, suggesting that catastrophic forgetting can occur.
For example, on the DCASE 2023 eval subset, Raw-SSL with BEATs achieves 63.60\%, whereas Dis-SSL-machine yields a substantially lower score of 53.78\%.
In contrast, Dis-SSL-PL preserves the performance of Raw-SSL and can further improve upon it.
Specifically, Dis-SSL-PL with BEATs achieves 65.37\% on the same subset and achieves the best average performance.

\subsubsection{Noise-robust feature transformation}
\label{sec:results_noise_robust}
\Cref{fig:dcase2025_hmean} shows the evaluation results for \ac{nrft}, where all labels other than the machine-type labels were concealed.
The results demonstrate the general effectiveness of \ac{nrft}, as it outperforms the standard Dis-multi-PL across different cluster ratios and \ac{ssl} models.
We can also see that Dis-multi-NRFT achieves performance comparable to that of Dis-multi-GT.
The results aggregated over the clean-data-available case show that smaller values of $K$ are more effective, indicating that restricting the feature space to a lower-dimensional clean-sound subspace effectively removes noise-related variations.
The results aggregated over the noise-data-available case show that \ac{nrft} is still effective with $K=D=64$ (i.e., without dimensionality reduction), indicating that noise covariance whitening itself contributes to suppressing the effects of noise on pseudo-label generation.
Although smaller values of $K$ tend to be more effective, performance improvements are obtained over a range of $K$ values and the effectiveness of NRFT is not limited to a specific choice of $K$. %

\Cref{fig:unified_machine_wise_umap_graph_dcase2025} shows machine-wise analyses of four machine types in the DCASE 2025 dataset.
Both the \ac{gevd}-based method using isolated noise samples (e.g., for Polisher) and the \ac{pca}-based method using clean machine sounds (e.g., for valve) improve performance.
In particular, for Polisher, the original BEATs feature space reflects both machine and noise characteristics, causing samples with the same attribute to be fragmented into multiple noise-dependent clusters.
\ac{nrft} successfully suppresses these noise-related variations and merges the fragmented clusters into attribute-wise clusters, resulting in a substantial performance improvement.
For valve, \ac{nrft} also reduces noise-related variations and improves performance, although some noise-dependent clusters remain.
For gearbox, \ac{nrft} suppresses noise-related variations and improves performance, but the resulting feature space still does not clearly reflect the attributes due to the limited representational capability of BEATs.
For AutoTrash, the original feature space already reflects the attributes to some extent, and \ac{nrft} does not degrade the performance.

A practical limitation of NRFT is its dependence on auxiliary recordings of clean machine sounds or isolated noise. Although we used only 100 auxiliary samples per machine type in our experiments, such recordings may be difficult to obtain in some practical settings. Future work will investigate how the amount of auxiliary data affects NRFT performance.

\section{Conclusion}
In this paper, we proposed pseudo-label distillation for discriminative \ac{asd}.
The proposed framework generates pseudo labels by clustering \ac{ssl} features and trains a compact discriminative frontend using the pseudo labels together with available machine-type labels and mixup.
We also proposed lightweight \ac{nrft} methods to improve the robustness of pseudo labels to noise variations in the training data.
\Ac{nrft} estimates a noise-robust feature transformation using a small amount of clean machine sounds or isolated noise samples.
Comprehensive evaluations on the DCASE 2020--2025 datasets using four \ac{ssl} models showed that pseudo-label distillation outperformed both the original \ac{ssl} models and discriminative models trained using only machine-type labels, while reducing the number of parameters and \acp{mac} to less than 10\% of those of the original \ac{ssl} models.
The analyses further showed that both machine-type labels and mixup contributed to performance improvements, that pseudo labels were also effective for fine-tuning \ac{ssl} models, and that the \ac{nrft} methods further improved performance.

\bibliographystyle{IEEEtran}%
\bibliography{bibfiles.bib}%

\begin{thebibliography}{10}
\providecommand{\url}[1]{#1}
\csname url@samestyle\endcsname
\providecommand{\newblock}{\relax}
\providecommand{\bibinfo}[2]{#2}
\providecommand{\BIBentrySTDinterwordspacing}{\spaceskip=0pt\relax}
\providecommand{\BIBentryALTinterwordstretchfactor}{4}
\providecommand{\BIBentryALTinterwordspacing}{\spaceskip=\fontdimen2\font plus
\BIBentryALTinterwordstretchfactor\fontdimen3\font minus
  \fontdimen4\font\relax}
\providecommand{\BIBforeignlanguage}[2]{{%
\expandafter\ifx\csname l@#1\endcsname\relax
\typeout{** WARNING: IEEEtran.bst: No hyphenation pattern has been}%
\typeout{** loaded for the language `#1'. Using the pattern for}%
\typeout{** the default language instead.}%
\else
\language=\csname l@#1\endcsname
\fi
#2}}
\providecommand{\BIBdecl}{\relax}
\BIBdecl

\bibitem{koizumi2020description}
Y.~Koizumi, Y.~Kawaguchi, K.~Imoto, T.~Nakamura, Y.~Nikaido, R.~Tanabe,
  H.~Purohit, K.~Suefusa, T.~Endo, M.~Yasuda, and N.~Harada, ``Description and
  discussion on {DCASE}2020 challenge task2: Unsupervised anomalous sound
  detection for machine condition monitoring,'' in \emph{Proc. DCASE}, 2020,
  pp. 81--85.

\bibitem{kawaguchi2021description}
Y.~Kawaguchi, K.~Imoto, Y.~Koizumi, N.~Harada, D.~Niizumi, K.~Dohi, R.~Tanabe,
  H.~Purohit, and T.~Endo, ``Description and discussion on {DCASE} 2021
  challenge task 2: Unsupervised anomalous detection for machine condition
  monitoring under domain shifted conditions,'' in \emph{Proc. DCASE}, 2021,
  pp. 186--190.

\bibitem{dohi2022description}
K.~Dohi, K.~Imoto, N.~Harada, D.~Niizumi, Y.~Koizumi, T.~Nishida, H.~Purohit,
  R.~Tanabe, T.~Endo, M.~Yamamoto, and Y.~Kawaguchi, ``Description and
  discussion on {DCASE} 2022 challenge task 2: Unsupervised anomalous sound
  detection for machine condition monitoring applying domain generalization
  techniques,'' in \emph{Proc. DCASE}, 2022, pp. 1--5.

\bibitem{dohi2023description}
K.~Dohi, K.~Imoto, N.~Harada, D.~Niizumi, Y.~Koizumi, T.~Nishida, H.~Purohit,
  R.~Tanabe, T.~Endo, and Y.~Kawaguchi, ``Description and discussion on {DCASE}
  2023 challenge task 2: First-shot unsupervised anomalous sound detection for
  machine condition monitoring,'' in \emph{Proc. DCASE}, 2023, pp. 31--35.

\bibitem{nishida2024description}
T.~Nishida, N.~Harada, D.~Niizumi, D.~Albertini, R.~Sannino, S.~Pradolini,
  F.~Augusti, K.~Imoto, K.~Dohi, H.~Purohit, T.~Endo, and Y.~Kawaguchi,
  ``Description and discussion on {DCASE} 2024 challenge task 2: First-shot
  unsupervised anomalous sound detection for machine condition monitoring,'' in
  \emph{Proc. DCASE}, 2024, pp. 111--115.

\bibitem{nishida2025description}
------, ``Description and discussion on {DCASE} 2025 challenge task 2:
  First-shot unsupervised anomalous sound detection for machine condition
  monitoring,'' in \emph{Proc. DCASE}, 2025, pp. 55--59.

\bibitem{fujimura2025asdkit}
T.~Fujimura, K.~Wilkinghoff, K.~Imoto, and T.~Toda, ``{ASDKit}: A toolkit for
  comprehensive evaluation of anomalous sound detection methods,'' in
  \emph{Proc. DCASE}, 2025, pp. 40--44.

\bibitem{lopez2020speaker}
J.~A. Lopez, H.~Lu, P.~Lopez{-}Meyer, L.~Nachman, G.~Stemmer, and J.~Huang, ``A
  speaker recognition approach to anomaly detection,'' in \emph{Proc. DCASE},
  2020, pp. 96--99.

\bibitem{giri2020self}
R.~Giri, S.~V. Tenneti, F.~Cheng, K.~Helwani, U.~Isik, and A.~Krishnaswamy,
  ``Self-supervised classification for detecting anomalous sounds,'' in
  \emph{Proc. DCASE}, 2020, pp. 46--50.

\bibitem{primus2020anomalous}
P.~Primus, V.~Haunschmid, P.~Praher, and G.~Widmer, ``Anomalous sound detection
  as a simple binary classification problem with careful selection of proxy
  outlier examples,'' in \emph{Proc. DCASE}, 2020, pp. 170--174.

\bibitem{wilkinghoff2023design}
K.~Wilkinghoff, ``Design choices for learning embeddings from auxiliary tasks
  for domain generalization in anomalous sound detection,'' in \emph{Proc.
  ICASSP}, 2023, pp. 1--5.

\bibitem{wilkinghoff2024self}
------, ``Self-supervised learning for anomalous sound detection,'' in
  \emph{Proc. ICASSP}.\hskip 1em plus 0.5em minus 0.4em\relax IEEE, 2024, pp.
  276--280.

\bibitem{kuroyanagi2025serial}
I.~Kuroyanagi, T.~Hayashi, K.~Takeda, and T.~Toda, ``{Serial-OE}: Anomalous
  sound detection based on serial method with outlier exposure capable of using
  small amounts of anomalous data for training,'' \emph{APSIPA Trans. Signal
  Inf. Process.}, vol.~14, no.~1, 2025.

\bibitem{liu2022anomalous}
Y.~Liu, J.~Guan, Q.~Zhu, and W.~Wang, ``Anomalous sound detection using
  spectral-temporal information fusion,'' in \emph{Proc. ICASSP}.\hskip 1em
  plus 0.5em minus 0.4em\relax IEEE, 2022, pp. 816--820.

\bibitem{saengthong2025deep}
P.~Saengthong and T.~Shinozaki, ``Deep generic representations for
  domain-generalized anomalous sound detection,'' in \emph{Proc. ICASSP}.\hskip
  1em plus 0.5em minus 0.4em\relax IEEE, 2025, pp. 1--5.

\bibitem{saengthong2026sub}
------, ``Sub-band spectral matching with localized score aggregation for
  robust anomalous sound detection,'' \emph{arXiv preprint arXiv:2603.13749},
  2026.

\bibitem{wilkinghoff2026temporal}
K.~Wilkinghoff, S.~Yadav, and Z.-H. Tan, ``Temporal pooling strategies for
  training-free anomalous sound detection with self-supervised audio
  embeddings,'' \emph{arXiv preprint arXiv:2603.04605}, 2026.

\bibitem{wilkinghoff2023on}
K.~Wilkinghoff and F.~Fritz, ``On using pre-trained embeddings for detecting
  anomalous sounds with limited training data,'' in \emph{Proc. EUSIPCO}, 2023,
  pp. 186--190.

\bibitem{jiang2024anopatch}
A.~Jiang, B.~Han, Z.~Lv, Y.~Deng, W.-Q. Zhang, X.~Chen, Y.~Qian, J.~Liu, and
  P.~Fan, ``{AnoPatch}: Towards better consistency in machine anomalous sound
  detection,'' in \emph{Proc. Interspeech}, 2024, pp. 107--111.

\bibitem{zheng2024improving}
X.~Zheng, A.~Jiang, B.~Han, Y.~Qian, P.~Fan, J.~Liu, and W.-Q. Zhang,
  ``Improving anomalous sound detection via low-rank adaptation fine-tuning of
  pre-trained audio models,'' in \emph{Proc. SLT}.\hskip 1em plus 0.5em minus
  0.4em\relax IEEE, 2024, pp. 969--974.

\bibitem{han2025exploring}
B.~Han, A.~Jiang, X.~Zheng, W.-Q. Zhang, J.~Liu, P.~Fan, and Y.~Qian,
  ``Exploring self-supervised audio models for generalized anomalous sound
  detection,'' \emph{IEEE Trans. Audio, Speech, Lang. Process.}, 2025.

\bibitem{fujimura2025improvements}
T.~Fujimura, I.~Kuroyanagi, and T.~Toda, ``Improvements of discriminative
  feature space training for anomalous sound detection in unlabeled
  conditions,'' in \emph{Proc. ICASSP}.\hskip 1em plus 0.5em minus 0.4em\relax
  IEEE, 2025, pp. 1--5.

\bibitem{wilkinghoff2023why}
K.~Wilkinghoff and F.~Kurth, ``Why do angular margin losses work well for
  semi-supervised anomalous sound detection?'' \emph{IEEE/ACM Trans. Audio,
  Speech, Lang. Process.}, 2023.

\bibitem{chen2023beats}
S.~Chen, Y.~Wu, C.~Wang, S.~Liu, D.~Tompkins, Z.~Chen, W.~Che, X.~Yu, and
  F.~Wei, ``{BEAT}s: Audio pre-training with acoustic tokenizers,'' in
  \emph{Proc. ICML}, vol. 202, 2023, pp. 5178--5193.

\bibitem{chen2024eat}
W.~Chen, Y.~Liang, Z.~Ma, Z.~Zheng, and X.~Chen, ``{EAT}: Self-supervised
  pre-training with efficient audio transformer,'' in \emph{Proc. IJCAI},
  K.~Larson, Ed., 2024, pp. 3807--3815.

\bibitem{dinkel2024scaling}
H.~Dinkel, Z.~Yan, Y.~Wang, J.~Zhang, Y.~Wang, and B.~Wang, ``Scaling up masked
  audio encoder learning for general audio classification,'' in \emph{Proc.
  Interspeech}, 2024.

\bibitem{fan2026fisher}
P.~Fan, A.~Jiang, S.~Zhang, X.~Zheng, Z.~Lv, B.~Han, W.~Liang, J.~Li, W.-Q.
  Zhang, Y.~Qian \emph{et~al.}, ``{FISHER}: A foundation model for multimodal
  industrial signal comprehensive representation,'' \emph{IEEE Trans. Ind.
  Inform.}, 2026.

\bibitem{niizumi2024masked}
D.~Niizumi, D.~Takeuchi, Y.~Ohishi, N.~Harada, and K.~Kashino, ``Masked
  {M}odeling {D}uo: Towards a universal audio pre-training framework,''
  \emph{IEEE/ACM Trans. Audio, Speech, Lang. Process.}, vol.~32, pp.
  2391--2406, 2024.

\bibitem{gong2022ssast}
Y.~Gong, C.-I. Lai, Y.-A. Chung, and J.~Glass, ``{SSAST}: Self-supervised audio
  spectrogram transformer,'' in \emph{Proc. AAAI}, 2022, pp. 10\,699--10\,709.

\bibitem{kuroyanagi2025improving}
I.~Kuroyanagi, T.~Fujimura, K.~Takeda, and T.~Toda, ``Improving anomalous sound
  detection through pseudo-anomalous set selection and pseudo-label utilization
  under unlabeled conditions,'' \emph{APSIPA Trans. Signal Inf. Process.},
  vol.~14, no.~1, pp. 1--28, 2025.

\bibitem{chen2026improving}
Z.~Chen, Y.~Zhang, and M.~Li, ``Improving anomalous sound detection with top-m
  pseudo-labeling,'' in \emph{Proc. Man-Mach. Speech Commun.}, J.~Jia, Z.~Wu,
  L.~Gao, G.~Huang, and Y.~Li, Eds., 2026, pp. 129--137.

\bibitem{fang2026improving}
X.~Fang, G.~Zhong, Q.~Wang, F.~Chu, L.~Wang, M.~Qian, M.~Cai, J.~Wu, J.~Gao,
  and J.~Du, ``Improving anomalous sound detection with attribute-aware
  representation from domain-adaptive pre-training,'' in \emph{Proc.
  ICASSP}.\hskip 1em plus 0.5em minus 0.4em\relax IEEE, 2026, pp.
  15\,892--15\,896.

\bibitem{deng2019arcface}
J.~Deng, J.~Guo, N.~Xue, and S.~Zafeiriou, ``{ArcFace}: Additive angular margin
  loss for deep face recognition,'' in \emph{Proc. CVPR}, 2019, pp. 4690--4699.

\bibitem{zhang2019adacos}
X.~Zhang, R.~Zhao, Y.~Qiao, X.~Wang, and H.~Li, ``{AdaCos}: Adaptively scaling
  cosine logits for effectively learning deep face representations,'' in
  \emph{Proc. CVPR}, 2019, pp. 10\,823--10\,832.

\bibitem{zhang2018mixup}
H.~Zhang, M.~Cisse, Y.~N. Dauphin, and D.~Lopez-Paz, ``mixup: Beyond empirical
  risk minimization,'' in \emph{Proc. ICLR}, 2018.

\bibitem{sandler2018mobilenetv2}
M.~Sandler, A.~Howard, M.~Zhu, A.~Zhmoginov, and L.-C. Chen, ``{MobileNetv2}:
  Inverted residuals and linear bottlenecks,'' in \emph{Proc. CVPR}, 2018, pp.
  4510--4520.

\bibitem{fujimura2025discriminative}
T.~Fujimura, I.~Kuroyanagi, and T.~Toda, ``Discriminative anomalous sound
  detection using pseudo labels, target signal enhancement, and ensemble
  feature extractors,'' in \emph{Proc. DCASE}, 2025, pp. 180--184.

\bibitem{zhang2026echo}
Y.~Zhang, J.~Liu, and M.~Li, ``{ECHO}: Frequency-aware hierarchical encoding
  for variable-length signals,'' in \emph{Proc. ICASSP}, 2026, pp. 4301--4305.

\bibitem{xie2020self}
Q.~Xie, M.-T. Luong, E.~Hovy, and Q.~V. Le, ``Self-training with noisy student
  improves {ImageNet} classification,'' in \emph{Proc. CVPR}, 2020, pp.
  10\,687--10\,698.

\bibitem{purohit2019mimii}
H.~Purohit, R.~Tanabe, T.~Ichige, T.~Endo, Y.~Nikaido, K.~Suefusa, and
  Y.~Kawaguchi, ``{MIMII Dataset}: Sound dataset for malfunctioning industrial
  machine investigation and inspection,'' in \emph{Proc. DCASE}, 2019, pp.
  209--213.

\bibitem{koizumi2019toyadmos}
Y.~Koizumi, S.~Saito, H.~Uematsu, N.~Harada, and K.~Imoto, ``{ToyADMOS}: A
  dataset of miniature-machine operating sounds for anomalous sound
  detection,'' in \emph{Proc. WASPAA}, 2019, pp. 308--312.

\bibitem{tanabe2021mimii}
R.~Tanabe, H.~Purohit, K.~Dohi, T.~Endo, Y.~Nikaido, T.~Nakamura, and
  Y.~Kawaguchi, ``{MIMII DUE}: Sound dataset for malfunctioning industrial
  machine investigation and inspection with domain shifts due to changes in
  operational and environmental conditions,'' in \emph{Proc. WASPAA}, 2021, pp.
  21--25.

\bibitem{harada2021toyadmos2}
N.~Harada, D.~Niizumi, D.~Takeuchi, Y.~Ohishi, M.~Yasuda, and S.~Saito,
  ``{ToyADMOS2}: Another dataset of miniature-machine operating sounds for
  anomalous sound detection under domain shift conditions,'' in \emph{Proc.
  DCASE}, 2021, pp. 1--5.

\bibitem{dohi2022mimii}
K.~Dohi, T.~Nishida, H.~Purohit, R.~Tanabe, T.~Endo, M.~Yamamoto, Y.~Nikaido,
  and Y.~Kawaguchi, ``{MIMII DG}: Sound dataset for malfunctioning industrial
  machine investigation and inspection for domain generalization task,'' in
  \emph{Proc. DCASE}, 2022.

\bibitem{harada2023toyadmos2}
N.~Harada, D.~Niizumi, D.~Takeuchi, Y.~Ohishi, and M.~Yasuda, ``{ToyADMOS2+}:
  New {ToyADMOS} data and benchmark results of the first-shot anomalous sound
  event detection baseline,'' in \emph{Proc. DCASE}, 2023, pp. 41--45.

\bibitem{niizumi2024toyadmos2}
D.~Niizumi, N.~Harada, Y.~Ohishi, D.~Takeuchi, and M.~Yasuda, ``{ToyADMOS2\#}:
  Yet another dataset for the {DCASE2024} challenge task 2 first-shot anomalous
  sound detection,'' in \emph{Proc. DCASE}, 2024, pp. 106--110.

\bibitem{albertini2024imad}
D.~Albertini, F.~Augusti, K.~Esmer, A.~Bernardini, and R.~Sannino, ``{IMAD-DS}:
  A dataset for industrial multi-sensor anomaly detection under domain shift
  conditions,'' in \emph{Proc. DCASE}, 2024, pp. 1--5.

\bibitem{harada2025toyadmos2025}
N.~Harada, D.~Niizumi, Y.~Ohishi, D.~Takeuchi, and M.~Yasuda, ``{ToyADMOS2025}:
  The evaluation dataset for the {DCASE2025T2} first-shot unsupervised
  anomalous sound detection for machine condition monitoring,'' in \emph{Proc.
  DCASE}, 2025, pp. 230--234.

\bibitem{wilkinghoff2025handling}
K.~Wilkinghoff, T.~Fujimura, K.~Imoto, J.~Le~Roux, Z.-H. Tan, and T.~Toda,
  ``Handling domain shifts for anomalous sound detection: A review of
  {DCASE}-related work,'' in \emph{Proc. DCASE}, 2025, pp. 20--24.

\bibitem{loshchilov2019decoupled}
I.~Loshchilov and F.~Hutter, ``Decoupled weight decay regularization,'' in
  \emph{Proc. ICLR}, 2019.

\bibitem{hu2022lora}
E.~J. Hu, Y.~Shen, P.~Wallis, Z.~Allen-Zhu, Y.~Li, S.~Wang, L.~Wang, and
  W.~Chen, ``Lo{RA}: Low-rank adaptation of large language models,'' in
  \emph{Proc. ICLR}, 2022.

\bibitem{okabe2018interspeech}
K.~Okabe, T.~Koshinaka, and K.~Shinoda, ``Attentive statistics pooling for deep
  speaker embedding,'' in \emph{Proc. Interspeech}, 2018, pp. 2252--2256.

\bibitem{chawla2002smote}
N.~V. Chawla, K.~W. Bowyer, L.~O. Hall, and W.~P. Kegelmeyer, ``{SMOTE}:
  synthetic minority over-sampling technique,'' \emph{J. Artif. Intell. Res.},
  vol.~16, pp. 321--357, 2002.

\bibitem{mcinnes2018umap}
L.~McInnes, J.~Healy, N.~Saul, and L.~Grossberger, ``{UMAP}: Uniform manifold
  approximation and projection,'' \emph{J. Open Source Softw.}, vol.~3, no.~29,
  2018, 63 pages.

\end{thebibliography}

\vspace{-40em}
\begin{IEEEbiographynophoto}{Takuya Fujimura} received an M.E. degree in informatics from Nagoya University, Nagoya, Japan, in 2024.
He is currently working toward a Ph.D. degree in informatics at Nagoya University.
His research interests include audio signal processing and machine learning.
He is a student member of the Acoustical Society of Japan (ASJ).
He received the 21st Best Student Presentation Award of ASJ in 2020 and the DCASE Challenge Task 2 Judges' Award in 2025 and 2026.
\end{IEEEbiographynophoto}
\vspace{-40em}
\begin{IEEEbiographynophoto}{Tomoki Toda} received his B.E. degree from Nagoya University, Nagoya, Japan, in 1999, and the M.E. and D.E. degrees from the Nara Institute of Science and Technology (NAIST), Ikoma, Japan, in 2001 and 2003, respectively. From 2003 to 2005, he was a Research Fellow with the Japan Society for the Promotion of Science. He was an Assistant Professor, from 2005 to 2011, and an Associate Professor, from 2011 to 2015, with NAIST. Since 2015, he has been a Professor with the Information Technology Center, Nagoya University. His research focuses on statistical approaches to speech and audio processing. He was the recipient of more than ten article/achievement awards, including the IEEE SPS 2009 Young Author Best Paper Award and the 2013 EURASIP-ISCA Best Paper Award (\textit{Speech Communication}).
\end{IEEEbiographynophoto}

\EOD

\end{document}